\documentclass[prl,twocolumn,superscriptaddress,floatfix,showpacs,amsmath,amssymb]{revtex4}
\usepackage{graphicx}
\begin{document}
\title{Drag Reduction by Polymer Additives in Decaying Turbulence}
\author{Chirag Kalelkar}
\email{kalelkar@physics.iisc.ernet.in}
\affiliation{Centre for Condensed Matter Theory, Department of Physics, Indian 
Institute of Science, Bangalore 560012, India.}
\author{Rama Govindarajan}
\email{rama@jncasr.ac.in}
\affiliation{Engineering Mechanics Unit, Jawaharlal Nehru Centre for Advanced 
Scientific Research, Jakkur, Bangalore 560064, India.}
\author{Rahul Pandit}
\email{rahul@physics.iisc.ernet.in}
\altaffiliation[\\ Also at ]{the Jawaharlal Nehru Centre for Advanced 
Scientific Research, Bangalore, India.}
\affiliation{Centre for Condensed Matter Theory,
Department of Physics, Indian Institute of Science, Bangalore 560012, India.}
\begin{abstract}
We present results from a systematic numerical study of decaying turbulence in 
a dilute polymer solution by using a shell-model version of the FENE-P 
equations. Our study leads to an appealing definition of drag reduction for 
the case of decaying turbulence. We 
exhibit several new results, such as the potential-energy spectrum of the 
polymer, hitherto unobserved features in the temporal evolution of the 
kinetic-energy spectrum, and characterize intermittency in such systems. 
We compare our results with the GOY shell model for fluid turbulence. 
\end{abstract}
\pacs{47.27.Gs, 83.60.Yz}
\maketitle
\begin{figure}
\includegraphics[height=1.8in]{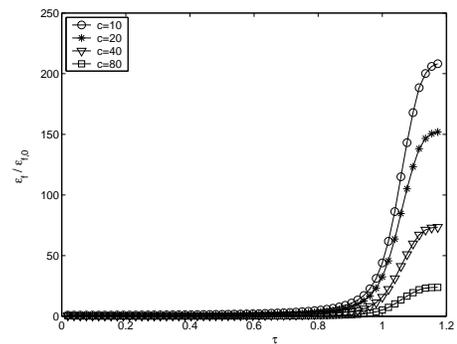}
\caption{\label{edissconc}The normalized kinetic energy dissipation rate 
${\mathcal E}_f/{\mathcal E}_{f,0}$ of the FENE-P shell model as a 
function of the dimensionless time $\tau$ for different values of 
concentration $c$, as 
specified in the legend.}
\end{figure}
\begin{figure}
\includegraphics[height=1.9in]{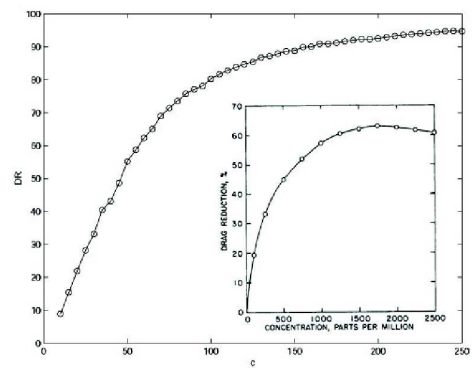}
\caption{\label{dragconc}Plot of the percentage drag reduction $DR$ (Eq. 
(\ref{dragreduction})) as a function of the concentration $c$. The inset shows 
a plot for the same quantities taken from Ref. \cite{Hoyt} (see the text for 
definitions).}
\end{figure}

The phenomenon of drag reduction by polymer additives\cite{Lumley}, 
whereby dilute solutions of linear, flexible, high-molecular-weight polymers 
exhibit frictional resistance to flow much 
lower than that of the pure solvent, has almost exclusively been studied 
within the 
context of {\it statistically steady} turbulent flows since the pioneering 
work of Toms\cite{Toms}. By contrast, there is an extreme scarcity of results 
concerning the effects of polymer additives on {\it decaying} 
turbulence\cite{Nadolink}. Experimental studies of decaying, homogeneous 
turbulence behind a grid indicate, for such dilute polymer solutions, a 
turbulent energy spectrum similar to that found without 
polymers\cite{Friehe,Mccomb}. However, flow visualization via 
die-injection tracers\cite{Mccomb} and 
particle image velocimetry\cite{Doorn} show an inhibition of small-scale 
structures in 
the presence of polymer additives. To the best of our knowledge 
{\it decaying} turbulence in such polymer 
solutions has not been studied numerically. We initiate such a study 
here by using a 
shell model that is well suited to examining the effects of polymer additives 
in turbulent flows that are homogeneous and in which bounding walls have no 
direct role. We obtain several interesting results including a {\it natural} 
definition of the percentage drag-reduction $DR$, which 
has been lacking for the case of decaying turbulence. We show that the 
dependence of $DR$ on 
the polymer concentration $c$ is in qualitative accord with 
experiments\cite{Lumley} as is the suppression of small-scale structures which 
we quantify by obtaining the filtered-wavenumber-dependence of the flatness of 
the velocity field.\\
We will use a shell-model version of the FENE-P 
(Finitely Extensible Nonlinear Elastic - Peterlin)\cite{Warner,Bird} 
model for dilute polymer solutions that has often been used for 
studying viscoelastic effects since it contains the basic characteristics of 
molecular stretching, orientation and finite extensibility seen in polymer 
molecules. A direct numerical 
simulation of the FENE-P equations is computationally prohibitive. This 
motivates the use of a 
shell model that captures the essential features of the FENE-P equations. 
Recent studies\cite{Benzi} have exploited 
a formal analogy\cite{Fouxon} of the FENE-P equations with those of 
magnetohydrodynamics (MHD) to construct such a shell model. We investigate 
decaying turbulence in a dilute polymer solution 
by developing a similar shell model for the FENE-P equations. 
The unforced FENE-P equations\cite{Warner,Bird} are
\begin{eqnarray}
\frac{\partial{\bf v}}{\partial{t}}+({\bf v\cdot\bigtriangledown}){\bf v}=
-\frac{\bigtriangledown p}{\rho_s}+\nu_s{\bigtriangledown^2\bf v}+
\bigtriangledown\cdot {\cal T},\\
\begin{split}
\frac{\partial{\cal R}_{\alpha\beta}}{\partial t}+({\bf v\cdot\bigtriangledown})
{\cal R}_{\alpha\beta}=&\frac{\partial v_\alpha}{\partial x_\gamma}{\cal R}_{\gamma\beta}+
\nonumber\\
&{\cal R}_{\alpha\gamma}\frac{\partial v_\beta}{\partial x_\gamma}-
\frac{1}{\nu_p}{\cal T}_{\alpha\beta},
\end{split}
\label{fenep}
\end{eqnarray}
where $p$ is the pressure, $\nu_s$ the kinematic viscosity of the solvent, 
 $\nu_p$ a `viscosity' parameter,  $\rho_s$ the density of the solvent, 
incompressibility is enforced via $\bigtriangledown\cdot{\textbf v}=0$, and  
the polymer conformation 
tensor is ${\cal R}_{\alpha\beta}\equiv\langle R_\alpha R_\beta\rangle/
R_0^2$, with the angular brackets indicating an average over 
polymer configurations, of the dyadic product of the end-to-end vector 
{\bf R}({\bf x},t) of the polymer molecules. The maximal extension of the 
polymer molecules is restricted by the condition 
$\langle R_{\gamma}^2\rangle<R_0^2$.
The contribution to the stress 
tensor because of the polymer is ${\cal T}_{\alpha\beta}=
\nu_p[P({\bf x},t){\cal R}_{\alpha\beta}-\delta_{\alpha\beta}]/\tau_p$, with 
$\delta_{\alpha\beta}$ the Kronecker delta, 
$\tau_p$ the time constant of the FENE-P model, and 
$P({\bf x},t)\equiv 1/(1-{\cal R}_{\gamma\gamma})$ (with repeated indices 
indicating a trace). 
The concentration of the polymer is parametrized here by $c\equiv 
\nu_p/\nu_s$.\\
Our shell-model version of the unforced FENE-P equations, obtained by 
generalising a shell model originally proposed for three-dimensional 
MHD\cite{Frick}, is
\begin{eqnarray}
\frac{d v_n}{d  t} &=&  \Phi_{n,vv} 
- \nu_s k^{2}_n v_n + \frac{\nu_p}{\tau_p}P(b)
\Phi_{n,bb} , \nonumber\\
\frac{d b_n}{d  t} &=&  \Phi_{n,vb} - 
\Phi_{n,bv} - \frac{1}{\tau_p}P(b) b_n,
\label{shell}
\end{eqnarray}
where $P(b)\equiv1/(1 - \sum_n |b_n|^2)$, $v_n$ and $b_n$ are 
complex, scalar variables representing the velocity 
and the (normalized) polymer end-to-end vector fields, respectively, 
with the discrete 
wavenumbers $k_n=k_02^n$ ($k_0$ sets the scale for 
wave-numbers), for shell index $n$ ($n=1\ldots N$, for $N$ shells), with 
$\Phi_{n,vv}=i(a_1k_nv_{n+1}v_{n+2}+a_2k_{n-1}v_{n+1}v_{n-1}+
a_3k_{n-2}v_{n-1}v_{n-2})$, $\Phi_{n,bb}=-i(a_1k_nb_{n+1}b_{n+2}+
a_2k_{n-1}b_{n+1}b_{n-1}+a_3k_{n-2}b_{n-1}b_{n-2})$, $\Phi_{n,vb}=
i(a_4k_nv_{n+1}b_{n+2}+a_5k_{n-1}v_{n-1}b_{n+1}+a_6k_{n-2}v_{n-1}b_{n-2})$, 
and $\Phi_{n,bv}=-i(a_4k_nb_{n+1}v_{n+2}+a_5k_{n-1}b_{n-1}v_{n+1}+
a_6k_{n-2}b_{n-1}v_{n-2})$.
As in Ref. \cite{Frick} we choose 
$a_1=1$, $a_2=a_3=-1/2$, $a_4=1/6$, $a_5=1/3$, and $a_6=-2/3$. 
We solve Eqs.~(\ref{shell}) numerically by using an Adams-Bashforth 
scheme\cite{Pisarenko} and double-precision arithmetic, with a step size 
$\delta t=10^{-2}$ and $N=22$ shells, 
with $k_0=1/16$ and $\nu_s=10^{-3}$. For all our runs (except 
those in FIGs. \ref{edissconc} and \ref{dragconc}) we set $c=100$. 
For numerical stability we add a nominal viscous term $-\nu_bk_n^2b_n$ to the 
shell-model equations for $b_n$ and set $\nu_b/\nu_s=10^{-13}$. With these  
parameter values, our code is stable for $1.0<\tau_p<7.8$, and we observe 
that the corresponding percentage drag reduction 
$DR$ (see below) lies in the range $63\%<DR<98\%$. For specificity we use 
$\tau_p=2.1$ for the data presented here. The initial velocity field is 
taken to be $v_n^0=k_n^{1/2}e^{i\theta_n}$ (for $n=1,2$), 
$v_n^0=k_n^{1/2}e^{-k_n^2}e^{i\theta_n}$ 
(for $3\le n\le N$) and the initial polymer field to be 
$b_n^0={k_n}^{1/2}e^{i\phi_n},$ 
with $\theta_n$ and $\phi_n$ independent random phases distributed 
uniformly between $0$ and $2\pi$.  In decaying turbulence, it is convenient 
to measure time in units of the initial large eddy-turnover time. For our 
shell model this is $\tau_0\equiv1/(v_{rms}^{0}k_1)$ with 
$v_{rms}^{0}\equiv[\langle\sum_n|v_n^0|^2\rangle]^{1/2}$, 
the root-mean-square value of the initial velocity (we find $\tau_0=5.2$). 
We use the dimensionless time 
$\tau\equiv t/\tau_0$ ($t$ is the product of the number 
of steps and $\delta t$). Our runs are ensemble averaged over $10^4$ 
independent initial conditions with different realizations of phases. We define 
$Re_v^{0}\equiv v^{0}_{rms}/(k_1\nu_s)$ to be the value of the initial Reynolds 
number (here $Re_v^{0}$ equals $12309$). Shell-model energy densities are 
defined as $E_a(k_n)\equiv \langle|a_n|^2/k_n\rangle$, with $a=v$ for the 
velocity 
field\cite{Frick,GOY} and $a=b$ for the polymer field. Equations (\ref{shell}) 
reduce to those for the Gledzer-Ohkitani-Yamada (GOY) shell 
model\cite{GOY} when the polymer-field terms are suppressed. For our 
GOY shell model runs, we use initial parameter values as for the 
FENE-P shell model to facilitate comparisons between the two.\\ 
\begin{figure}
\includegraphics[height=1.8in]{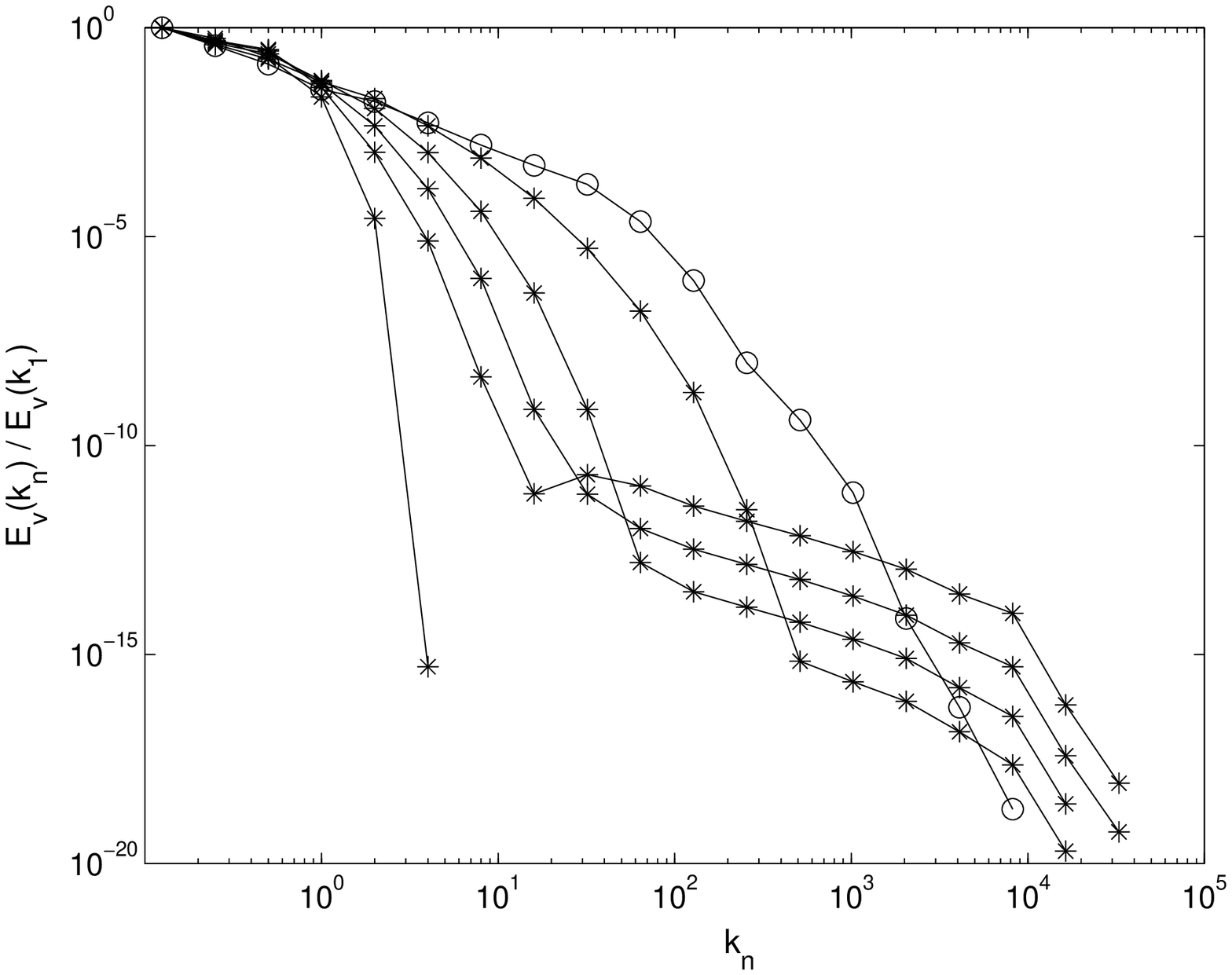}
\includegraphics[height=1.8in]{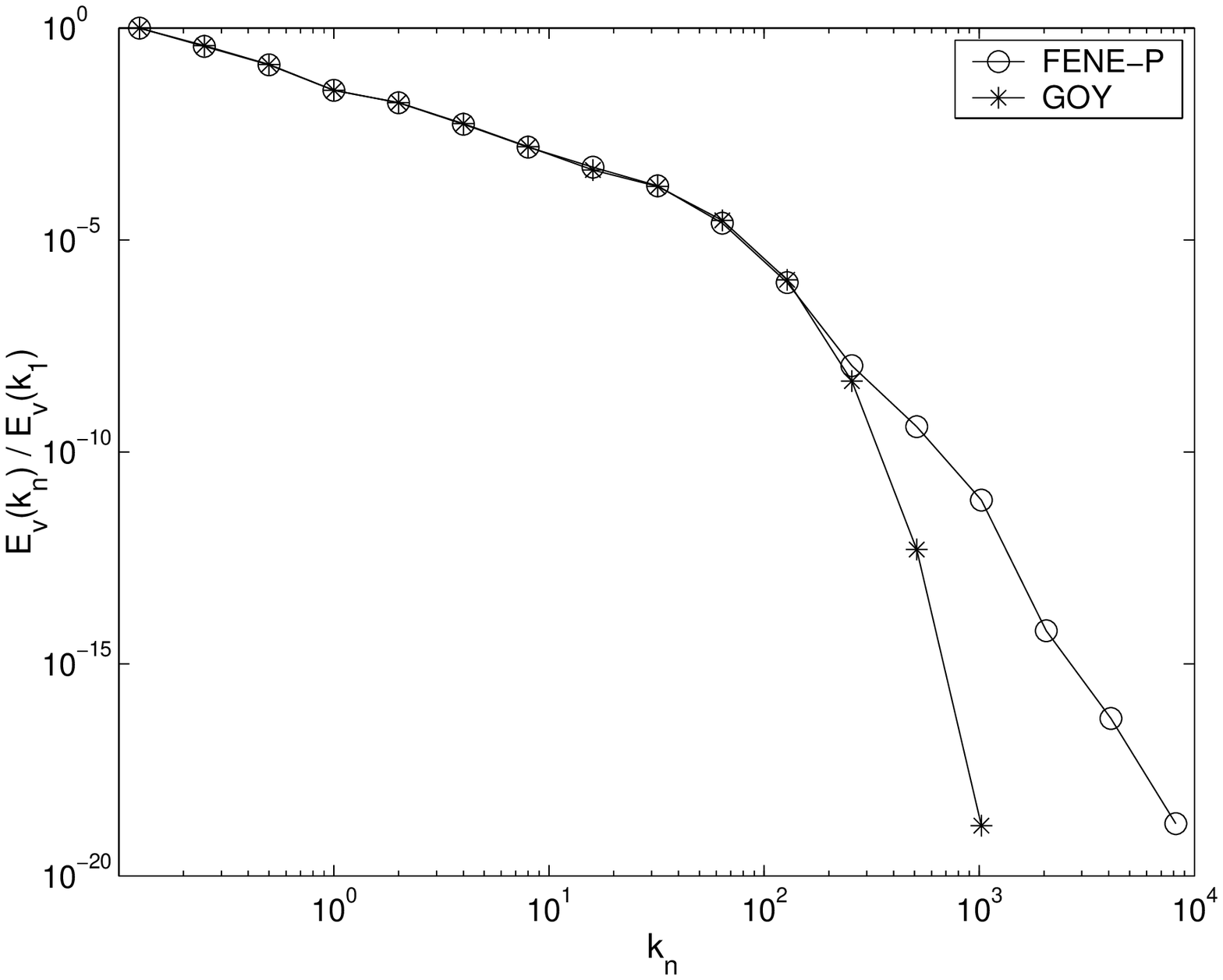}
\caption{\label{evol-especv}(a) Log-log plots of the temporal evolution of the 
normalized kinetic energy spectra $E_v(k_n)/E_v(k_1)$ of the FENE-P shell 
model as a function of the wavenumber $k_n$. The plot with open circles 
is calculated at cascade completion.\\(b) Log-log plots of the normalized 
kinetic energy spectra $E_v(k_n)/E_v(k_1)$ as a function of the 
wavenumber $k_n$ for the FENE-P and the GOY shell models at cascade 
completion. The observed slope is $-1.67\pm0.01$ for the range $0.25<k_n<64$.}
\end{figure}
Figure \ref{evol-especv}(a) shows the time evolution of the normalized kinetic 
energy spectrum $E_v(k_n)/E_v(k_1)$ (successive curves separated by time 
intervals of $0.2\tau$). We see a cascade of the energy to large 
wavenumbers after which the shape of the spectrum does not change appreciably 
but the energy decays. We observe the evolution of a flat portion in 
the spectrum that vanishes upon cascade completion (plot with open circles). 
Figure \ref{evol-especv}(b) compares 
kinetic-energy spectra at cascade completion for our model 
(Eqs. (\ref{shell})) and for the GOY shell model. In the inertial range, both 
spectra are indistinguishable and show a Kolmogorov-type $k^{-5/3}$ 
behavior with an observed slope of $-1.67\pm0.01$ (with errors from 
least-square fits), a result consistent with experiments\cite{Friehe,Mccomb} of 
decaying, 
homogeneous turbulence behind a grid for a dilute polymer solution. However, 
significant differences show up in the 
dissipation range: 
the spectrum for the FENE-P shell model falls much more slowly than its 
GOY-model counterpart indicating greatly reduced dissipation at large 
wavenumbers. Experimental\cite{Friehe,Mccomb} energy spectra do not cover 
as large a 
range of spatial scales as we can cover in our shell-model study, and thus, to 
the best of our knowledge, these dissipation-range discrepancies of the energy 
spectra, with and without polymer additives, have not been noticed earlier.
We note that our results in FIG. \ref{evol-especv}(b) distinctly differ from 
corresponding results\cite{Benzi} for statistically steady turbulence, 
where a tilt in 
the spectrum has been observed at low wavenumbers in the FENE-P shell model 
relative to that obtained from the GOY shell model.\\ 
\begin{figure}
\includegraphics[height=1.9in]{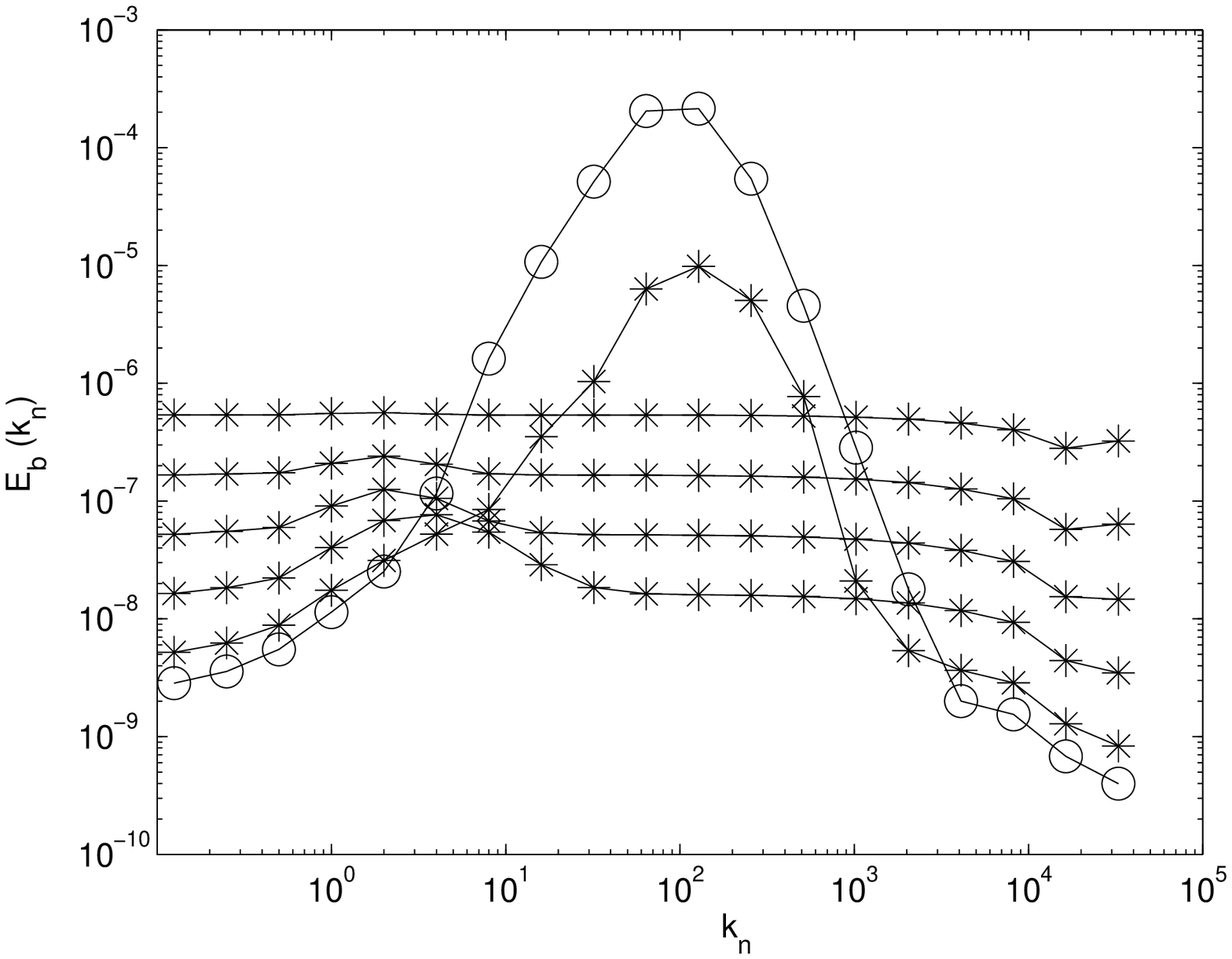}
\includegraphics[height=1.9in]{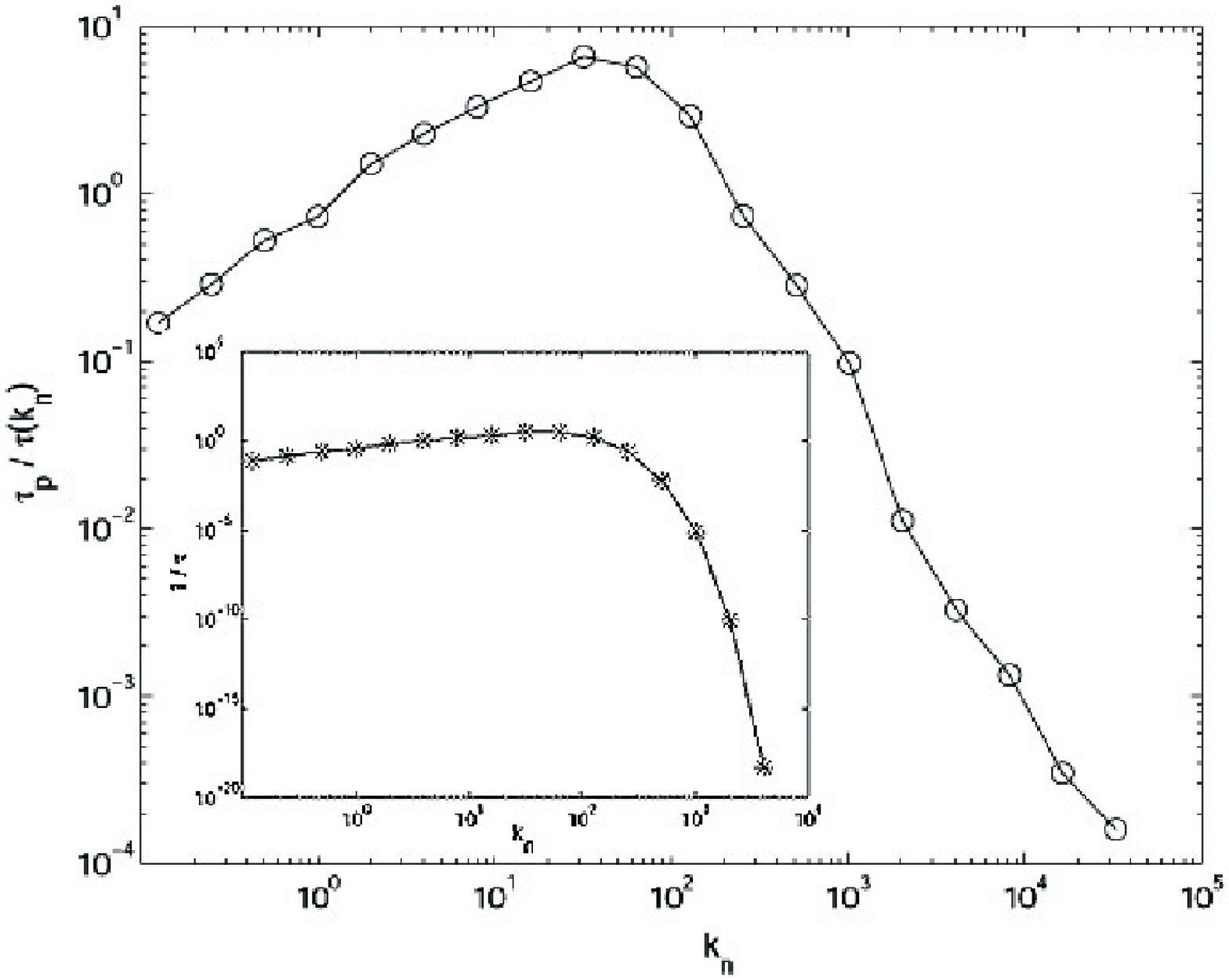}
\caption{\label{especb}(a) Log-log plots of the temporal evolution of the 
potential energy spectrum $E_b(k_n)$ of the polymer field as a function of the 
wavenumber $k_n$. The plot with open circles is calculated at cascade 
completion.\\(b) Log-log plot of the ratio of the time constant 
$\tau_p$ of the FENE-P shell model and the turbulence 
time-scale $\tau(k_n)$ as a function of the wavenumber $k_n$ at cascade 
completion. The inset shows the inverse of the turbulence time-scale 
$\tau(k_n)$ as a function of the wavenumber $k_n$ for the GOY shell model at 
cascade completion.} 
\end{figure}
In FIG. \ref{especb}(a), we display the time evolution of the potential-energy 
spectrum of the polymer $E_b(k_n)$ (with a temporal separation of $0.2\tau$). 
Starting from an initially flat spectrum, we 
observe the appearance and subsequent growth of a protuberance that bulges out 
maximally on cascade completion (plot with open circles) at a 
wavenumber corresponding to the value, of order unity, of the ratio of the 
polymer time constant $\tau_p$ and the turbulence time scale 
$\tau(k_n)\equiv 1/(k_n|v(k_n)|)$ (FIG. \ref{especb}(b)). The result is 
in agreement with a hypothesis (for statistically steady turbulence) in 
Ref. \cite{Degennes} wherein a polymer molecule, immersed in an eddy with a 
turbulent time-scale comparable to the polymer relaxation time, undergoes 
a `coil-stretch' 
transition with an increment in the potential-energy spectrum at 
the wavenumber corresponding to the inverse of the eddy size. The inset 
in FIG. \ref{especb}(b) is a plot of the inverse of the turbulence time-scale
 $\tau(k_n)$ as a function of the wavenumber $k_n$ for the GOY shell model. 
In both plots, within the inertial range, $\tau(k_n)\sim k_n^{-2/3}$, a result 
consistent with the $-5/3$ power-law in the kinetic energy spectrum.\\ 
A log-log plot of the normalized kinetic 
energy dissipation rate ${\mathcal E}_f/{\mathcal E}_{f,0}$ of the FENE-P 
shell model versus dimensionless time $\tau$ for different values of $c$ is 
shown in FIG. 
\ref{edissconc} (with ${\cal E}_f/{\cal E}_{f,0}\equiv\langle
\sum_nk_n^2|v_n|^2\rangle/\langle\sum_nk_n^2|v_n^0|^2\rangle$, the additional 
index $0$ indicating values calculated at initial times). 
The reduction in the peak value with respect 
to the value at initial times, with increasing concentration, is indicative of 
an enhanced value of the stored elastic potential energy in the polymer 
molecules due to their extension. The analogous plot for the GOY shell model 
is identical to the plot for $c=10$ in FIG. \ref{edissconc}. 
We are therefore led to the following natural definition of the 
percentage drag reduction $DR$ for decaying turbulence: 
\begin{eqnarray}
DR\equiv\Big(\frac{{\cal E}_{g,m}/{\cal E}_{g,0}-{\cal E}_{f,m}/{\cal E}_{f,0}}{{\cal E}_{g,m}/{\cal E}_{g,0}}\Big)\times 100,
\label{dragreduction}
\end{eqnarray}
where the kinetic energy dissipation rates 
${\cal E}_a\equiv\langle\sum_nk_n^2|v_n|^2\rangle$ (the subscript $a=f$ for 
the FENE-P shell model and $a=g$ for the GOY shell model) are calculated 
upon cascade completion when the dissipation rate is a maximum (indicated by 
an additional subscript $m$) and normalized by 
their values at initial times (indicated by an additional subscript $0$). 
With the choice of initial parameter values as specified above, 
${\mathcal E}_{g,m}/{\cal E}_{g,0}$ equals $234.96\pm0.01$.\\
In FIG. \ref{dragconc}, we use Eq. (\ref{dragreduction}) to plot $DR$ as a 
function of $c$. The inset 
figure from Ref. \cite{Hoyt} is a similar plot for a dilute solution of 
Carrageenan\cite{fn} (a seaweed derivative) in a pipe-flow Reynolds 
number of $14000$. 
The qualitative 
agreement with a laboratory experiment (for statistically steady 
turbulence) supports our definition of drag reduction for 
decaying turbulence.\\ 
\begin{figure}
\includegraphics[height=1.7in]{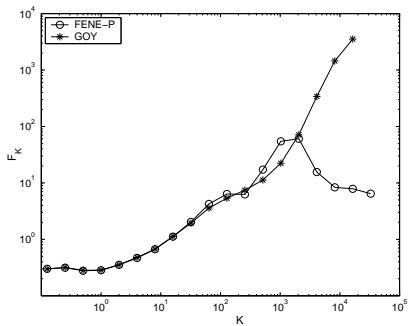}
\caption{\label{flat}Plot of the flatness $F_K$ as a function of the 
filtered wavenumber $K$ (see the text for definitions) at cascade completion 
for the FENE-P and the GOY shell models.}
\end{figure}
Laboratory experiments in both statistically steady\cite{Dam} and 
decaying\cite{Mccomb,Doorn} turbulent flows of 
dilute polymer solutions show an inhibition of small-scale structures and 
narrower probability distribution functions of velocity differences\cite{Dam}. 
We plot in FIG. \ref{flat}, the 
flatness $F_K\equiv\langle |v^{>}_K|^4\rangle/\langle |v^{>}_K|^2\rangle^2$ 
($\langle|v^{>}_K|^4\rangle\equiv\langle\sum_n|v_n|^4\rangle$, 
$\langle|v^{>}_K|^2\rangle\equiv\langle\sum_n|v_n|^2\rangle$, with 
$n=K...N$), $K=1...N$, as a function of the `filtered'\cite{Frisch} 
wavenumber $K$ for the FENE-P and the GOY shell models at cascade completion. 
We observe 
that, in the GOY shell model, the flatness $F_K$ exhibits 
unbounded growth for large wavenumbers, an indication of strong intermittency 
in the dissipation scales. However, for the FENE-P shell model, we observe that 
the flatness is greatly reduced relative to that for GOY and, in fact, 
{\it decreases} in the dissipation scales. Our results are consistent, 
therefore, with laboratory experiments which show a suppression of small 
structures that 
would imply reduced intermittency in the dissipation range of our shell 
model.\\ 
Laboratory experiments\cite{Mccomb,Doorn} of decaying turbulence behind 
grids indicate a reduced decay rate of the kinetic energy in a dilute polymer 
solution, relative to the pure solvent. In the initial period of 
decay, before the 
integral scale of turbulence becomes of the order of the size of the system 
(the minimum wavenumber, in the case of shell models), we observe a decay 
rate of $-1.80\pm0.01$ for the FENE-P shell model and a decay rate of 
$-2.01\pm0.02$ for the GOY shell model 
(a result consistent with Ref. \cite{Lohse}).\\
In conclusion, then, we have presented results from a systematic 
numerical study of 
decaying turbulence in a dilute polymer solution by employing a shell-model 
for the FENE-P equations. This leads to a natural definition of drag reduction 
for such a system and new results on the potential- and kinetic energy 
spectra which are in qualitative agreement with experimental findings.
\begin{acknowledgments}
C.K. thanks Tejas Kalelkar for useful discussions and CSIR (India) for 
financial support; R.P. thanks the 
Indo-French Centre for Promotion of 
Advanced Scientific Research (IFCPAR Project No. 2404-2) and the Department of 
Science and Technology (India) Grant to the Centre for Condensed Matter 
Theory (IISc).
\end{acknowledgments}

\end{document}